\documentclass{llncs}

\usepackage{url,amsmath,amssymb,a4wide}

\newcommand{\imp}{\rightarrow}
\newcommand{\biimp}{\leftrightarrow}
\newcommand{\all}{\forall}
\newcommand{\ex}{\exists}

\newcommand{\nec}{\Box} 
\newcommand{\pos}{\Diamond} 
\newcommand{\ess}[2]{#1 \ \mathit{ess.} \ #2}
\newcommand{\NE}{\mathit{NE}}

\title{Formalization, Mechanization and Automation 
  of \\ G\"{o}del's Proof of God's Existence\thanks{This work has been
    supported by the German Research Foundation under grant
    BE2501/9-1.}}

\author{
  Christoph Benzm\"{u}ller\inst{1} 
  \and 
  Bruno Woltzenlogel Paleo\inst{2}
}

\authorrunning{C.\~Benzm\"{u}ller \and B.\~Woltzenlogel Paleo}

\institute{
  Dahlem Center for Intelligent Systems, Freie Universit\"{a}t Berlin, Germany\\
  \email{c.benzmueller@gmail.com}
  \and 
  Theory and Logic Group, Vienna University of Technology, Austria \\
  \email{bruno@logic.at}
}

\begin{document}

\maketitle

\bigskip

\noindent
\textbf{Update (31/08/2017):} The abstract below, uploaded to arXiv on
21/08/2013, was the first communication of the computer-assisted
formalization of G\"odel's ontological proof. Since then, the
following longer papers have been published:
\cite{W50,J28,B15,C39,C40,C41,C42,C44,C46,C52,W55,J32,C55,C60,B16,B66,J35,J36,C65}.

\bigskip
\bigskip

Attempts to prove the existence (or non-existence) of God by means of
abstract ontological arguments are an old tradition in philosophy and
theology.  G\"{o}del's proof \cite{Goedel1970,GoedelNotes} is a modern culmination of
this tradition, following particularly the footsteps of Leibniz.
G\"{o}del defines God as a being who possesses all \emph{positive} properties.
He does not extensively discuss what positive properties are, 
but instead he states a few reasonable (but debatable) axioms that they should satisfy.
Various slightly different versions of axioms and definitions have
been considered by G\"{o}del and by several philosophers who commented
on his proof
(cf. \cite{sobel2004logic,AndersonGettings,Fitting,Adams,ContemporaryBibliography}). 

Dana Scott's version of G\"odel's proof \cite{ScottNotes} employs the
following axioms (\textbf{A}), definitions (\textbf{D}), corollaries
(\textbf{C}) and theorems (\textbf{T}), and it proceeds in the
following order:\footnote{ A1, A2, A5, D1, D3 are logically
  equivalent to, respectively, axioms 2, 5 and 4 and definitions 1 and
  3 in G\"odel's notes \cite{Goedel1970,GoedelNotes}. 
  A3 was introduced by Scott \cite{ScottNotes} and 
  could be derived from G\"odel's axiom 1 and
  D1 in a logic with infinitary conjunction. 
  A4 is a weaker form of G\"odel's axiom 3. 
  D2 has an extra conjunct $\phi(x)$ lacking in G\"odel's definition 2; 
  this is believed to have been 
  an oversight by G\"odel \cite{Hazen}.}

\allowdisplaybreaks[1] 
\begin{description}
\item[A1] Either a property or its negation is positive, but not
  both:  \hfill 
  $\all \phi [P(\neg \phi) \biimp \neg P(\phi)]$
\item[A2] A property necessarily implied \\ by a
  positive property is positive: \phantom{b} \hfill 
  $\all \phi \all \psi [(P(\phi) \wedge \nec \all x [\phi(x)
  \imp \psi(x)]) \imp P(\psi)]$ 
\item[T1] Positive properties are possibly exemplified: \hfill $\all \varphi [P(\varphi) \imp \pos \ex x \varphi(x)]$
\item[D1] A \emph{God-like} being possesses all positive properties: \hfill
  $G(x) \biimp \forall \phi [P(\phi) \to \phi(x)]$
\item[A3]  The property of being God-like is positive: \hfill   $P(G)$
\item[C\phantom{1}] Possibly, God exists: \hfill $\pos \ex x G(x)$
\item[A4]  Positive properties are necessarily positive: \hfill 
  $\all \phi [P(\phi) \to \Box \; P(\phi)]$
\item[D2] An \emph{essence} of an individual is  \\ a property possessed by it and \\ necessarily implying any of its properties:
  \phantom{b} \hfill $\ess{\phi}{x} \biimp \phi(x) \wedge \all
  \psi (\psi(x) \imp \nec \all y (\phi(y) \imp \psi(y)))$
\item[T2]  Being God-like is an essence of any
  God-like being: \hfill $\all x [G(x) \imp \ess{G}{x}]$ 
\item[D3] \emph{Necessary existence} of an individual is \\ the necessary exemplification of all its essences: 
  \phantom{b} \hfill $\NE(x) \biimp \all \phi [\ess{\phi}{x} \imp \nec \ex y \phi(y)]$
\item[A5] Necessary existence is a positive property: \hfill $P(\NE)$
\item[T3] Necessarily, God exists: \hfill $\nec \ex x G(x)$ 
\end{description}

\noindent
Scott's version of G\"{o}del's proof has now been 
analysed for the first-time
with an unprecedent degree of detail 
and formality with the help of theorem provers; cf.~\cite{FormalTheologyRepository,ComputationalPhilosophyRepository}. 
The following has been done (and in this order):
\begin{itemize}
\item A detailed natural deduction proof.
\item A formalization of the axioms, definitions and theorems in the TPTP THF syntax \cite{J22}.
\item Automatic verification of the consistency of the axioms and 
definitions with Nitpick \cite{Nitpick}.
\item Automatic demonstration of the theorems with the provers LEO-II \cite{LEO-II} and Satallax \cite{Satallax}.

\item A step-by-step formalization using the Coq proof assistant \cite{Coq}.

\item A formalization using the Isabelle proof assistant \cite{Isabelle}, where the theorems (and some additional lemmata) have been automated with Sledgehammer \cite{Sledgehammer} and Metis \cite{Hurd03first-orderproof}.
\end{itemize}

G\"{o}del's proof is challenging to formalize and verify because it
requires an expressive logical language with modal operators
(\emph{possibly} and \emph{necessarily}) and with
quantifiers for individuals and properties.  Our computer-assisted formalizations rely on an
embedding of the modal logic into classical higher-order logic with
Henkin semantics \cite{J23,B9}. The formalization is thus essentially
done in classical higher-order logic where  quantified modal logic is
emulated.

In our ongoing computer-assisted study of G\"odel's proof we have
obtained the following results:
\begin{itemize}
\item The basic modal logic K is sufficient for proving T1, C and T2. 
\item Modal logic S5 is not needed for proving T3; the logic KB is
  sufficient. 
\item Without the first conjunct $\phi(x)$ in D2 the set of axioms 
  and definitions would be inconsistent.
\item For proving theorem T1, only the left to right direction of
  axiom A1 is needed. However, the backward direction of A1 is
  required for proving T2.

\end{itemize}

This work attests the maturity of contemporary interactive and
automated deduction tools for classical higher-order logic and
demonstrates the elegance and practical relevance of the
embeddings-based approach.  Most importantly, our work opens new
perspectives for a computer-assisted theoretical philosophy.  The
critical discussion of the underlying concepts, definitions and axioms
remains a human responsibility, but the computer can assist in
building and checking rigorously correct logical arguments. In case of
logico-philosophical disputes, the computer can check the disputing
arguments and partially fulfill Leibniz' dictum: Calculemus --- Let us
calculate!

\end{document}